\documentclass{PoS}
\newcommand\be{\begin{equation}}
\newcommand\ee{\end{equation}}

\title{The role of multigrid algorithms for LQCD}

\ShortTitle{The role of multigrid algorithms for LQCD}

\author{Ronald Babich\\
Center for Computational Science, Boston University, 
  Boston, MA 02215\\
   E-mail: \email{rbabich@bu.edu}}
\author{James Brannick\\
Department of Mathematics, The Pennsylvania State University,
    University Park, PA 16802\\
   E-mail: \email{brannick@psu.edu}}
\author{\speaker{Richard C. Brower}%
\\
 Department of Physics, Boston University,
        Boston, MA 02215\\
        E-mail: \email{brower@bu.edu}}
\author{Michael A. Clark\\
 Harvard-Smithsonian Center for Astrophysics and \\
Initiative in Innovative Computing,  Harvard University,
          Cambridge,  MA 02138\\
        E-mail: \email{mikec@seas.harvard.edu}}
\author{Saul D.  Cohen\\
    Center for Computational Science, Boston University, Boston,
  MA 02215\\
        E-mail: \email{sdcohen@bu.edu}}
\author{James C. Osborn\\
Argonne Leadership Computing Facility,
Argonne, IL 60439 \\
        E-mail: \email{osborn@alcf.anl.gov}}
\author{Claudio Rebbi\\
 Department of Physics, Boston University,
        Boston, MA 02215\\
        E-mail: \email{rebbi@bu.edu}}

      \abstract{ We report on the first successful QCD multigrid
        algorithm which demonstrates  constant convergence rates
        independent of quark mass and lattice volume for the Wilson
        Dirac operator. The new ingredient is the adaptive method for
        constructing the near null space on which the coarse grid
        multigrid Dirac operator acts.  In addition we speculate on
        future prospects for extending this algorithm to the Domain
        Wall and Staggered discretizations, its exceptional
        suitability for high performance GPU code and its potential
        impact on simulations at the physical pion mass.}

\FullConference{The XXVII International Symposium on Lattice Field Theory\\
		 July 26-31, 2009\\
		 Peking University, Beijing, China}

\begin{document}

\section{Introduction}
Perhaps the most severe computational challenge facing  lattice 
Quantum Chromodynamics is the divergent  cost as one
approaches  the chiral limit required for the experimental
value of the pion mass.  Similar difficulties confront other
strongly-coupled field theories
conjectured for physics beyond the standard model. The cause
is well known.  As the quark  mass, $m_q$, approaches zero,
the lattice Dirac operator becomes singular, $|\lambda_{min}|\;
\rightarrow 0$, causing ``critical slowing down'' of the 
iterative solvers typically used to find the propagators. This is
unavoidable for all local ``unigrid'' solvers.

It has been almost 20 years since the first
attempts~\cite{Kalkreuter:1994fz} were made  to
apply recursive multigrid preconditioning to the Dirac operator in
lattice QCD. The basic idea of using a coarse representation of the
Dirac operator on lattices of increasing lattice spacing appears at
first to be an obvious extension of the basic principles central to
the renormalization group itself. Indeed early attempts, generally
inspired by this observation, did succeed in formulating a variety of
gauge invariant coarsening schemes but they all failed to improve
convergence at the length scale $l_{\sigma}$, where the underlying
lattice gauge field becomes rough. Indeed as Brower, Edwards, Rebbi
and Vicari~\cite{Brower:1991xv} demonstrated this failure occurred
uniformly when the product of the mass gap $m_q$ and the coherence
length $l_\sigma$ is of order one: $m_q l_\sigma = O(1)$. Apparently
this failure occurs when the ``renormalization'' is highly
non-perturbative.

Recently the application of a new adaptive procedure~\cite{Brezina:2004} has decisively broken this
barrier eliminating ``critical slowing down'' as the mass gap goes to
zero~\cite{Brannick:2007ue}. Here we give a heuristic introduction to this
breakthrough and speculate on further developments.

\section{Adaptive Multigrid}

While the detailed design of an appropriate adaptive multigrid
algorithm for QCD requires considerable effort as reported in
Ref.~\cite{Brannick:2007ue} the underlying concept is rather
straightforward.  We seek to accelerate the solver for a differential
operator discretized on a hypercubic lattice with spacing $a$
\be
D_{xy}\psi_{y} = b_x \; ,
\ee
by preconditioning it with a coarse operator $\hat D$ at a larger
lattice spacing $\hat a > a$.  To be concrete our example is the Wilson
lattice Dirac operator,
\be
D_{xy}(U) =  (4 + m)\delta_{xy} +  
\sum^4_{\mu=1}[\frac{\gamma_\mu -1}{2} U_\mu(x)\delta_{x,y+\mu} 
- \frac{\gamma_\mu + 1}{2}  U^\dagger_\mu(x) \delta_{x+\mu,y}] \; ,
\ee
where we have suppressed the indices for the 3x3
(dense) SU(3) color matrices $U^{ab}_\mu(x)$ and the 4x4 (sparse)
spinor matrices $\gamma^{ij}_\mu$ in the tensor product.  The fine
Dirac matrix, $D$, operates on a complex vector space $V$ of dimension $12
L^3\times T$.

\begin{figure}[h!]
\begin{center}
\includegraphics[width=0.8\textwidth]{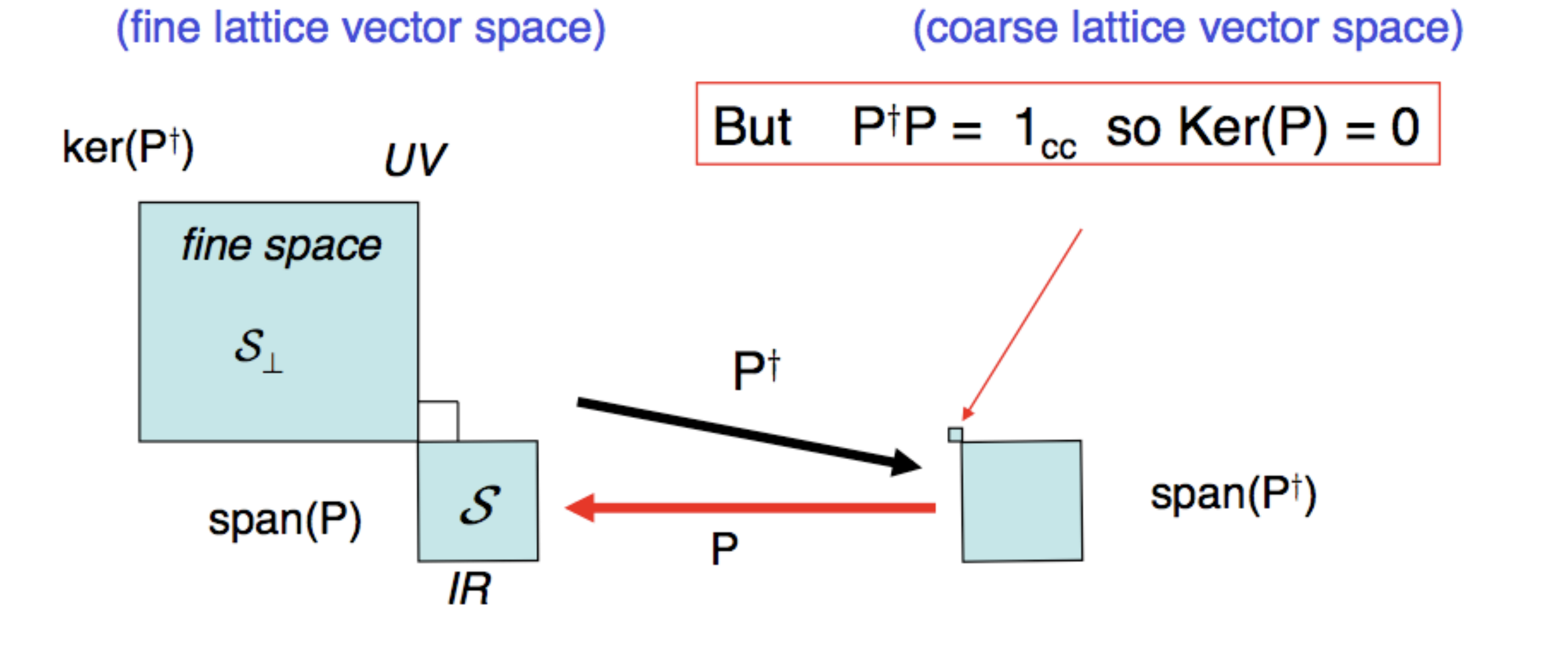}
\caption{The non-square prolongation matrix P with $ker(P) = 0$
  defines a one to one map $P:\hat V = S$ from the coarse vector space
  $\hat V$ into the near null subspace $S$ of the fine vector space:
  $V = S + S_\perp$.  The fundamental theorem  of linear algebra gives $S =
  span(P)$, $\hat V = span(P^\dagger)$ and $rank(P) = rank(P^\dagger)
  = dim(S)$.}
\label{fig:Proj}
\end{center}
\end{figure}

Critical slowing down is caused by eigenvectors with small
eigenvalues. This offending subspace is the near null space of our
operator: $D: S \simeq 0$. Multigrid methods require us to split the
fine vectors space $V$ into this near null space $S$ and its
orthogonal complement $S_\perp$: $V = S + S_\perp$, in the language of
the renormalization group, splitting the IR (near null) from the UV
(rough) modes. One may view this splitting as a generalization of
red/black or Schwartz block decompositions and the resultant
preconditioning matrix as akin to using the Schur compliment. This
splitting is achieved by a non-square prolongation matrix $P$ which
maps the coarse space into the near null space S,
\be
P: \hat V \rightarrow S \; ,
\ee
as illustrated in Fig~\ref{fig:Proj}. 
Then the multigrid cycle constructs a coarse matrix, $\hat D =
R D P$, as the product of the prolongator ($P$) to the near null
space on the fine lattice, the fine operator ($D$) and  a restriction
operator ($R$)  back to the
coarse lattice.   We use the Galerkin form  by setting $R = P^\dagger$.

To understand  intuitively how one constructs this mapping,
consider multigrid for the classic example of a d-dimensional
discretized Laplace operator.  The near null eigenvectors are
literally smooth, dominated by low Fourier components. An obvious
interpolation consists of piecewise constant functions on regular
blocks to define the coarse degrees of freedom. For example on each
$4^d$ block labeled by $\hat x$ we many introduce the prolongator (or
interpolating matrix), 
\be
P_{x  \hat x} = \frac{\theta_{\hat x}(x)}{2^d} \quad, \quad
\theta_{\hat x}(x) = 
\left\{
\begin{array}{ll}
1  & x \in \hat x \; \mbox{block} \\
0  & x \in\!\!\!\!\!\!\!/\; \; \hat x \; \mbox{block} 
\end{array}
\right. \; ,
\label{eq:Poperator}
\ee
where the blocking ``theta function'' , $ \theta_{\hat x}(x)$, is 1 (true) for x
inside and 0 (false) outside the block $\hat x$.  The normalization is
chosen so that $[P^\dagger P]_{\hat x \hat y} = \delta_{\hat x, \hat
  y}$. The span of this space consists of all linear combinations of 
these basis vectors: $\psi_x = \sum_{\hat x}
c_{\hat x} P_{x \hat x}$.  Solving the coarse problem exactly
for the error would reduce the residue to $ r' = {\cal P}r = (1 - D P
\frac{1}{P^\dagger D P} P^\dagger) r$, where
\be
{\cal P} = 1 -  D P \frac{1}{P^\dagger D P} P^\dagger \quad, \quad
{\cal P}^2 = {\cal P}
\ee
is the Petrov-Galerkin (oblique) projection operator with eigenvalues
0 and 1. This projector completely removes the near null space from
the residue: $P {\cal P} = 0$ but the transverse space $S_\perp$
of rough modes are left intact. To damp them out a smoother on the
fine lattice must also be applied.

Fortunately this basic construction carriers over to the non-trivial
example of lattice QCD. However to construct a parameterization
for the coarse lattice Dirac operator, a piecewise constant interpolation
is entirely inappropriate because of the almost random background
gauge matrices $U$ connecting nearest neighbor sites.  The insight of the
adaptive approach is to use the slow convergence of near null
components itself to define through the Galerkin scheme the coarse
operator.  One starts with a random fine vector 
and attempts to solve the homogeneous equation,
\be 
D(U) \psi^{(\alpha)} \simeq 0 \; ,
\ee
for an element at critical mass. After a few iterations this yields a
global near null vector, which is subsequently broken into blocks 
as in Eq.~\ref{eq:Poperator} and
used to construct a trial multigrid scheme. Then if this putative
scheme is slow to converge one uses it to solve again for a new near null vector
and repeats until a set of near null vectors, $(\psi^{(1)},
\psi^{(2)}, \cdots, \psi^{(N_\nu)})$ is found that eliminates critical
slowing down. The prolongator is therefore given by restricting each
global vector to blocks by $\theta_{\hat x}(x)\psi^{(\alpha)}_x$ and
orthonormalizing the basis on each bock to define the near null
subspace $S$,
\be
P_{x;  \hat x,\alpha} = \; \mbox{orthonormal basis for} \;
\{\theta_{\hat x}(x)\psi^{(\alpha)}_x \}  \; .
\ee
Again the near null space is spanned by this basis: $\psi_x =
\sum_{\hat x ,\alpha} c_{\hat x, \alpha} P_{x; \hat x,\alpha}$.  The
precise form of the adaptive iteration, the minimum number of global
near null vectors $N_\nu$ and the blocking configuration are all devised to find an efficient
multigrid preconditioner with minimal complexity. The contrast with
earlier attempts to construct multigrid algorithms for QCD appears to
be rather small. In the projective multigrid
scheme~\cite{Brower:1991xv}, near null vectors were found block by
block imposing Dirichlet boundary condition, more like a Schwarz
method. Basically by reversing the procedure to first finding global
near null vectors and second restricting them to blocks we have the
adaptive multigrid approach. This is typical of multigrid methods that
simple changes have profound consequences. The devil is in the
details.

\section{Performance of MG for Wilson Dirac Operator}

\begin{center}
\begin{figure}
 \includegraphics[width = 0.5\textwidth]{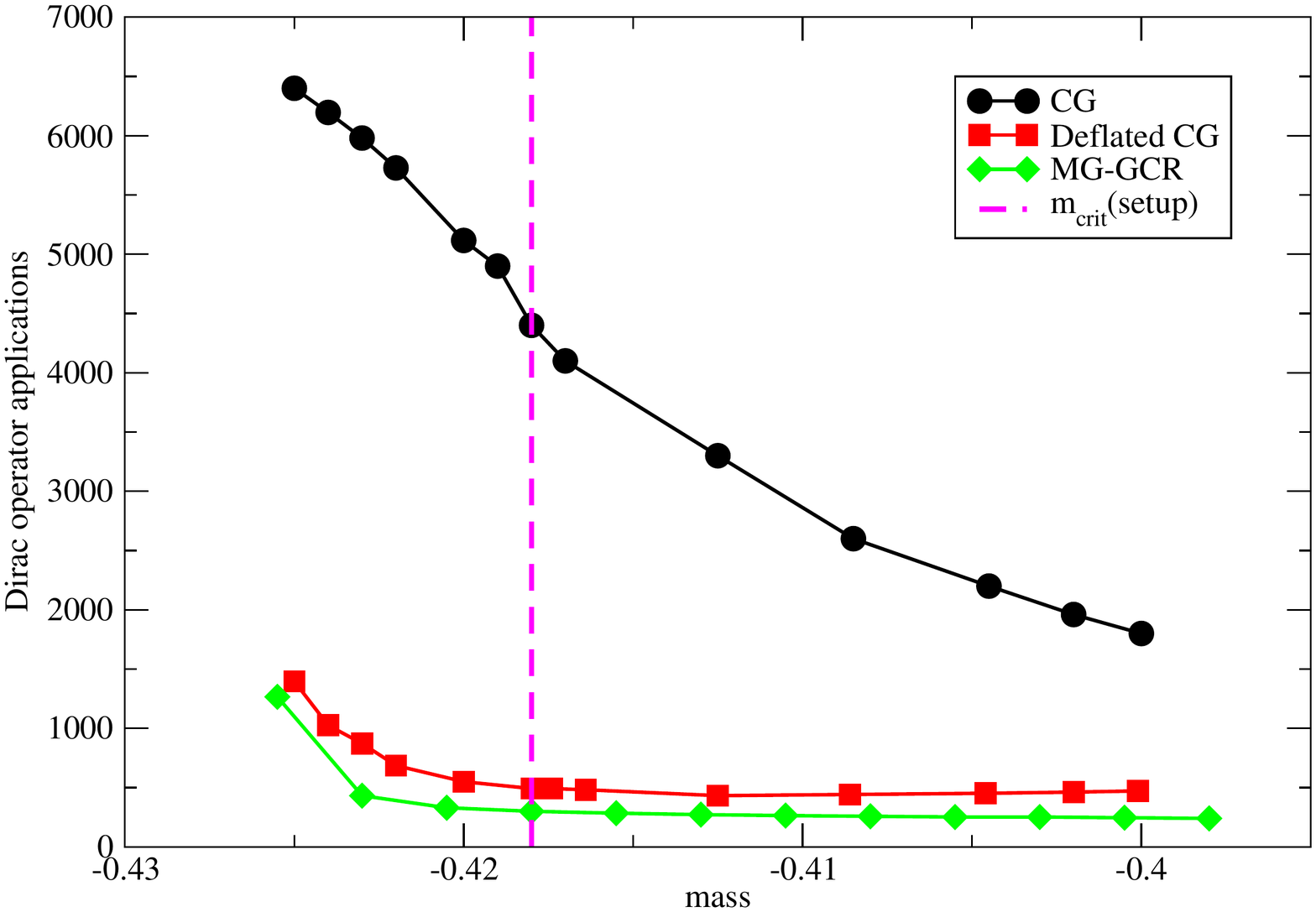}
  \includegraphics[width = 0.5\textwidth]{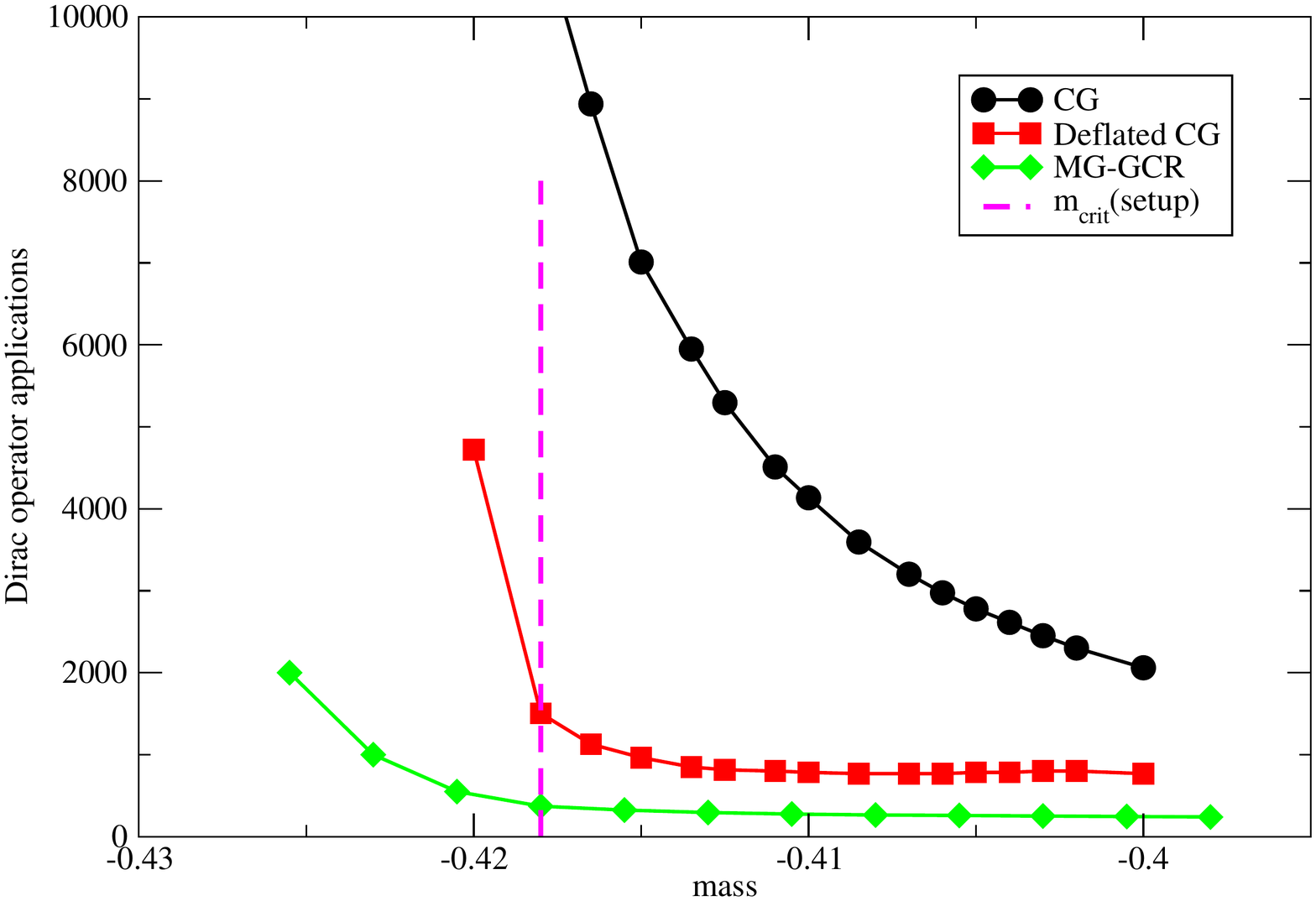}
  \caption{Comparison of CG, deflated CG\cite{Stathopoulos:2007zi} and MG-GCR total number of Wilson matrix-vector operations until convergence (point sources, \(V=16^3\times64\) (left plot), \(V=24^3\times64\) (right plot), \(\beta = 5.5\), \(m_{crit} = -0.4175\), \(N_v = 20\) (MG-GCR), \(N_v=240\) (deflated CG), solver tolerance \(=10^{-8}|b|\)). }
  \label{fig:4d}
\end{figure}
\end{center}

\begin{figure}[h!]
\begin{minipage}[c]{0.45\textwidth}
\includegraphics[width=1.0\textwidth]{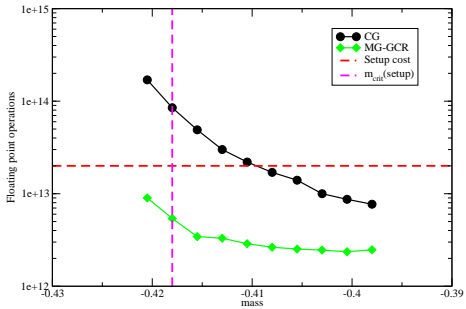}
\caption{Comparison of adaptive MG algorithm with the
conventional red/black preconditioned CG algorithm on $32^3\times 96$ lattice. }
\label{fig:MGqcd}
\end{minipage}
\hskip 1 cm
\begin{minipage}[c]{0.43\textwidth}
\begin{tabular}{|c|c|c|c|}
\hline
Mass: & $16^3 \times 64$ & $24^3 \times 64$ & $32^3 \times 96$ \\
\hline \hline
-.3980	&	40 &   40  &  41  \\
\hline
-.4005   &     41   & 41  &	 42 \\
\hline
-.4030 	  &    42  &   42  &	 43 \\
\hline
-.4055 	 &     42  &  43   &	 43 \\ 
\hline
-.4080    &     43 &   44  &	 45 \\ 
\hline
-.4105    &     44 &    46  &	 49 \\ 
\hline
-.4130 	 & 45  &  49  &	 52  \\  
\hline
-.4155 	&    47 &   54	& 57 \\ 
\hline
\end{tabular}
\vskip 0.8 cm
{ \small {\bf Table 1:} Fine grid iteration count as function of lattice
size and quark mass.}
\end{minipage}
\end{figure}

There are many technical details that are critical to an efficient
adaptive multigrid algorithm for the Wilson Dirac matrix.  Experience
first guided us to coarsen all color and Dirac degrees of freedom on
$4^4$ space-time blocks.  However for the Wilson Dirac operator, which
is neither Hermitian or normal, it proved to be important to preserve
the special property of $\gamma_5$-Hermiticity, $D^\dagger = \gamma_5
D \gamma_5$ on the coarse level by splitting each block into two
sub-blocks for $\gamma_5 = \pm 1$ labeled by $\sigma_3 = \pm 1$ so that
$\sigma_3 P = P \gamma_5$.  Finally we implemented a 3 level W-cycle
MG algorithm with 4 post smoothing iterations, using a GCR(8) outer
Krylov solver on the finest
level and a  CG complete solve on the normal equations on the coarsest. We
have clearly achieved a successful MG algorithm for the Wilson
operator which shows little or no sign of critical slowing down as
function of the quark mass or lattice size. Already it is competitive
with EigCG deflation~\cite{Stathopoulos:2007zi} on rather modest
lattice sizes (see Figs.~\ref{fig:4d})  and it will become increasingly
superior as the lattice become larger since the complexity of exact
deflation scale like $O(|V|^2)$ whereas multigrid scales 
no worse than $O(|V| log |V|)$ where $|V|$ is the volume of the lattice
or size of the fine vectors space.
With $N_\nu = 20$ trial near null vectors this is a very successful
multigrid method as illustrated in Fig.~\ref{fig:MGqcd}.  The
horizontal line is the set up cost of constructing the multigrid
operator.  Table 1 shows that the iteration count is nearly
independent of lattice size and the quark mass, down to the physical
pion mass ($m= - .4155$).

We are still at the beginning of additional improvements. For example
we have recently combined the multigrid algorithm with red/black
preconditioning yielding an additional 30\% improvement and we are
experimenting with exposing the full Dirac spin structure (not just
the chiral structure) on the coarser blocks.  It should be noted that
our choices were guided to a degree by physical intuition based on
chiral symmetry and the 't Hooft null states associate with isolated
instantons but to date there is no precise physical understanding or
rigorous mathematical analysis to explain the success of multigrid
QCD.  Further experimentation and more refined applied mathematical
tools are needed to approach an optimal method.

\section{Future directions}

Let us turn to future directions we are pursuing with the caveat that
until we have constructed and benchmarked these extensions, the
improvements are speculations based on our current experience. First we
have begun to design algorithms for both Staggered and Domain Wall
fermion discretizations. For Staggered fermions the technical barrier
appears modest, since the operator is normal and anti-Hermitian in the
chiral limit. However the ``species doubling'' quadruples the size of the near
null space and the Asqtad or HISQ improvements increase the potential
complexity of the coarsening. Still with the much larger lattices in
production we expect to find a very useful implementation.  For the
Domain Wall the technical issues are much more subtle but a strategy
is emerging.  The operator is not only non-Hermitian but the eigenvalues
do not have positive real parts as was the case for Wilson and
Staggered fermions. Indeed it is essentially a 5-d Wilson operator
with the wrong sign mass. However the potential advantage of Domain Wall
multigrid is greater.  The 5 dimensional Domain Wall matrix
operates in a larger vector space and is less well conditioned
because of the heavy flavor modes in the 5th dimension, but its near
null space is still four dimensional. Thus the truncation to the
coarse lattice is more dramatic and in principle there is more to be
gained in a multigrid algorithm. Similar remarks hold for the 
overlap formulation of the Dirac operator but the outer iteration
has the advantage of being a normal matrix with positive real mass gap.

A complete suite of multigrid algorithms for Staggered, Wilson and
chiral fermion actions holds out the promise of a major reduction in the
cost of Dirac inverters for the analysis stage of lattice QCD
ensembles.  As the physics correlators for lattice QCD have expanded in
the USQCD collaborations the relative number of flops devoted to
analysis is now exceeding 50\%.  In addition the multigrid kernel can
be used in a variance reduction strategy for stochastic estimators of
disconnected quark diagrams~\cite{Babich:2007jg}.We  are also  beginning to develop inverters compliant with the
SciDAC API for general distribution. Firsts we are extending
the API to accommodate the multiple lattices and to
implement  the interpolation and prolongation operators. Also
we are optimizing the $N_\nu \times N_\nu$ complex matrix operations
needed for the coarse operators.  These algorithms will
be freely distributed on the SciDAC software webpages.

In  principle  MG  inverters  can  be implemented  in  HMC  codes  for
generating  lattice  ensembles  as  well.  A  critical  step  in  this
application is to amortize the set up cost of constructing the coarse
operators as  the gauge fields  evolve in molecular dynamics  time. In
this regard L\"uscher has demonstrated~\cite{Luscher:2007es} that the
subspace update for his  ``little Dirac'' operator (which is essentially
equivalent to  our first  coarse operator $\hat  D$) can be  used for
several HMC  time steps combined  with a chronological  procedure for
incremental change  in the near  null space. This strongly  suggests that
the construction of the multigrid inverter is not a serious overhead.
Efficient parallel code is of course another requirement.
\begin{figure}[h!]
\begin{minipage}[c]{0.5\textwidth}
\includegraphics[width = 1.0\textwidth]{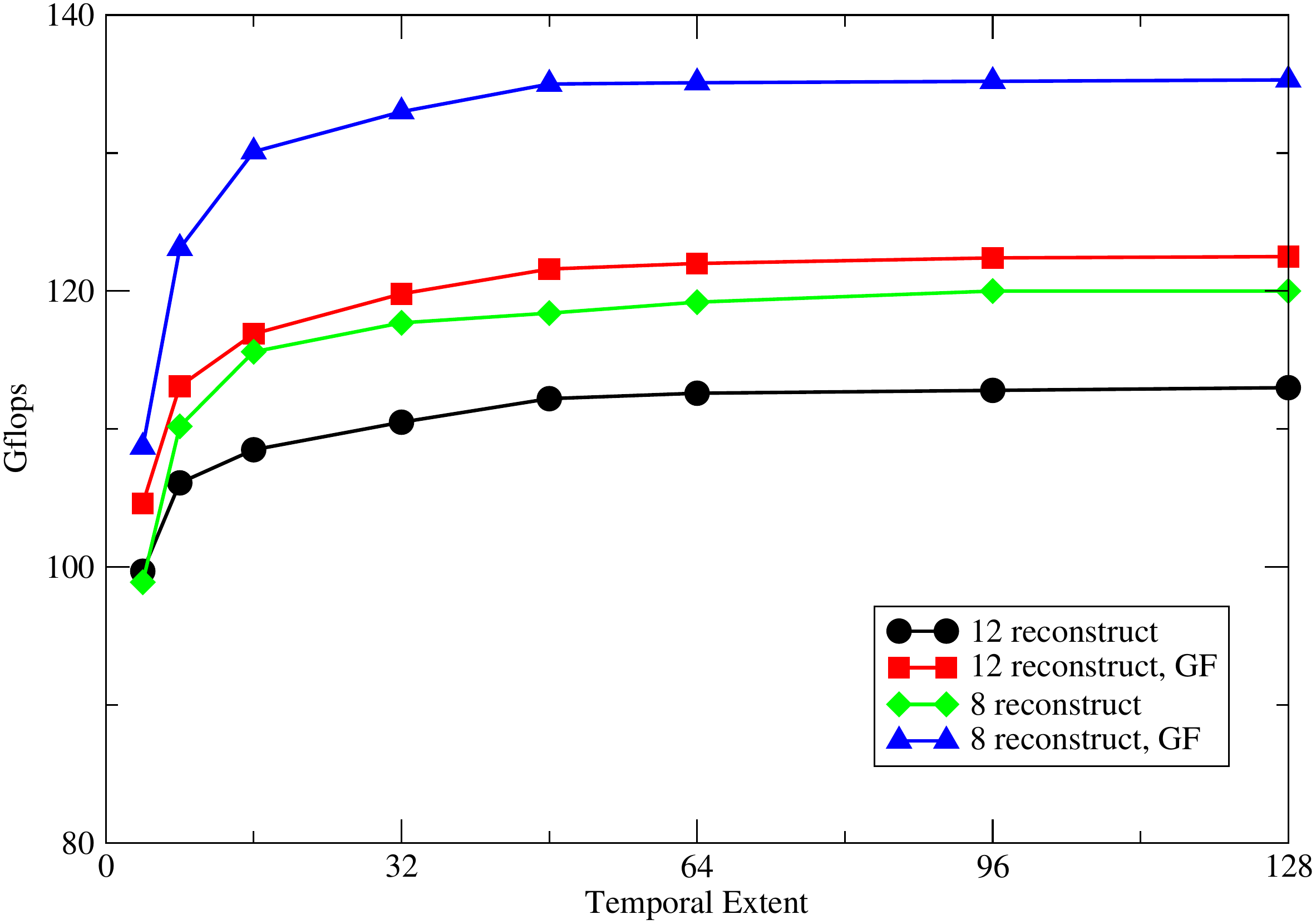}
 \caption{\label{fig:wilson-single-pc-flops}Performance of 
   single precision even-odd preconditioned Wilson-Dirac matrix-vector
   product on a GTX 280 ~\cite{Clark:2009wm}. GF denotes temporal
   gauge fixing (lattice volume = \(24^3 \times \mbox{Temporal
     Extent}\)).}
\end{minipage}
\hskip 1 cm
\begin{minipage}[c]{0.43\textwidth}
\begin{tabular}{|l|c|c|c|}\hline
Kernel & Kernel & CG  & BiCGstab  \\ 
Prec. & (Gflops) &  (Gflops)   &  (Gflops)   \\ \hline
Half 12  & 202.2 & 170.6 & 152.5 \\ \hline
SP 8  & 134.1 & 110.1 & 105.1 \\ \hline
SP 12  & 122.1 & 102.4 & 98.6 \\ \hline
DP 12  & 35.4 & 33.5 & 29.3 \\ \hline
\end{tabular}
\vskip 1.2 cm
{\small {\bf Table 2:} Performance comparison of the
  matrix-vector kernels with the associated CG and BiCGstab solvers
  on the GeForce GTX 280 (lattice volume =
  \(24^3\times48\))~\cite{Clark:2009wm}.
Gauge field stored as 12 or 8 floats.}
\end{minipage}
\label{fig:GPUperf}
\end{figure}

Finally it is worth closing with a comment on GPU computing.  At
Boston University we have implemented a highly efficient Wilson Dirac
CG and BiCGstab implementations for the Nvidia GPU written in the CUDA
extension of C~\cite{Clark:2009wm}. This gives roughly a 5x advantage
in cost performance for Dirac inversions (see
Fig.~\ref{fig:GPUperf}). 
Preliminary analysis
indicates that in many ways this architecture is well suited for the
multigrid inverter discussed above. The full implementation of this in
efficient code has begun and if successful promises a multiplicative
advantage in cost per Dirac inverter as the product of hardware and
algorithm advances. Even without extensions to the generation of
lattices the combined effect of GPU and MG has the potential of
dropping the cost of analysis of these lattices by several orders of
magnitude relative to current practices.

\begin{acknowledgments}
This work was supported in part by US DOE
grants DE-FG02-91ER40676 and DE-FC02-06ER41440 and NSF grants DGE-0221680,
PHY-0427646,  OCI-0749300 and  OCI-0749202.
\end{acknowledgments}


\begin{thebibliography}{99}

\bibitem{Kalkreuter:1994fz} For a review of earlier work see:
  T.~Kalkreuter,
  ``Multigrid methods for propagators in lattice gauge theories,''
  J.\ Comput.\ Appl.\ Math.\  {\bf 63}, 57 (1995)
  [arXiv:hep-lat/9409008].

\bibitem{Brower:1991xv}
  R.~C.~Brower, R.~G.~Edwards, C.~Rebbi and E.~Vicari,
  ``Projective multigrid for Wilson fermions,''
  Nucl.\ Phys.\  B {\bf 366}, 689 (1991).

\bibitem{Brezina:2004}
M.~Brezina, R.~Falgout, S.~MacLachlan, T.~Manteuffel, S.~McCormick, and
  J.~Ruge.
\newblock Adaptive smoothed aggregation ($\alpha${SA}).
\newblock {\em Siam J. Sci. Comput.}, 25:1896--1920, 2004.





\bibitem{Brannick:2007ue}
  J.~Brannick, R.~C.~Brower, M.~A.~Clark, J.~C.~Osborn and C.~Rebbi,
  ``Adaptive Multigrid Algorithm for Lattice QCD,''
  Phys.\ Rev.\ Lett.\  {\bf 100} (2008) 041601
  [arXiv:0707.4018 [hep-lat]].
  
\bibitem{Stathopoulos:2007zi}
  A.~Stathopoulos and K.~Orginos,
  ``Computing and deflating eigenvalues while solving multiple right hand side
  linear systems in Quantum Chromodynamics,''
  arXiv:0707.0131 [hep-lat].




\bibitem{Babich:2007jg}
  R.~Babich, R.~Brower, M.~Clark, G.~Fleming, J.~Osborn and C.~Rebbi,
  ``Strange quark contribution to nucleon form factors,''
  [arXiv:0710.5536 [hep-lat]].


\bibitem{Luscher:2007es}
  M.~Luscher,
  ``Deflation acceleration of lattice QCD simulations,''
  JHEP {\bf 0712}, 011 (2007)
  [arXiv:0710.5417 [hep-lat]].


\bibitem{Clark:2009wm}
  M.~A.~Clark, R.~Babich, K.~Barros, R.~C.~Brower and C.~Rebbi,
  ``Solving Lattice QCD systems of equations using mixed precision solvers on
  GPUs,''
  arXiv:0911.3191 [hep-lat].


\end{thebibliography}
\end{document}